# Highly-efficient, diffraction-limited laser emission from a Vertical External Cavity Surface-emitting Organic Laser

Hadi Rabbani-Haghighi, Sébastien Forget, Sébastien Chénais*, Alain Siove

*Laboratoire de Physique des Lasers, Université Paris 13 / CNRS, 93430 Villetaneuse, France*

*Corresponding author: sebastien.chenais@univ-paris13.fr

We report on a solid-state laser structure being the organic counterpart of the Vertical External-Cavity Surface-Emitting Laser (VECSEL) design. The gain medium is a poly (methyl methacrylate) film doped with Rhodamine 640, spin-casted onto the High-Reflectivity mirror of a plano-concave resonator. Upon pumping by 7-ns pulses at 532 nm, a diffraction-limited beam (M²=1) was obtained, with a conversion efficiency of 43%; higher peak powers (2kW) could be attained when resorting to shorter (0.5 ns) pump pulses. The spectrum was controlled by the thickness of the active layer playing the role of an intracavity etalon; tunability is demonstrated over up to 20 nm.

Organic solid-state lasers offer the potential of low-cost, compact and broadly tunable coherent sources over the whole visible spectrum [1]. They may find applications in the fields of, *e.g.*, polymer optical fiber telecommunications [2], bio/chemo-sensing [3] or spectroscopy [4]. In the perspective of achieving direct [5] or indirect [6] electrical pumping of organic semiconductors, most of recent studies have been dealing with low-loss resonator structures such as distributed feedback resonators [7, 8] or microcavities [9, 10]. In the meantime, less attention has been paid to the question of optimizing conversion efficiency, beam quality and power scaling capability. Low-threshold operation and high-output power are indeed incompatible statements in a single device: whereas weak output coupling and small active volume are desirable for low threshold lasing, achieving high output power leads to an optimal transmission for the output coupler [11] and requires keeping the pump fluence to a low level to cope with photo-bleaching and thermal issues. Furthermore, low-loss compact resonators do not generally provide diffraction-limited beams [12].

A combination of high efficiency, high output power and high beam quality is achievable with external resonators. A typical external-cavity organic laser consists of a thick (~ mm to cm) dye-doped polymeric sample prepared by *e.g.* radical polymerization [13] or sol-gel technique [14], which has to be polished to optical quality prior to its insertion in an open stable resonator. Preparation of such samples is complex and not compatible with organic semiconductor processing techniques. As a simpler alternative, it has been proposed to spin-coat an organic film directly onto one of the mirrors [15-17]. However pumping with femtosecond pulses not only limited the cavity length to a few µm, making this type of device very similar in nature to microcavity lasers, but also led to modest efficiencies (1.7%) [17].

In this Letter we report on a Vertical External Cavity Surface-emitting Organic Laser (VECSOL, upon analogy with the VECSEL structure [18]). It consists of an organic PMMA film doped with Rhodamine 640, spin-casted onto the High-Reflectivity mirror of a plano-concave resonator, pumped in the nanosecond regime. It shares the same general properties as VECSELs: diffraction-limited output, high efficiency, power scaling capability upon a simple increase of pump spot size (hence without higher photodegradation), possibility of long (cm) cavities enabling the insertion of intracavity elements such as spatial or spectral filters. It is compatible with organic semiconductors and hence with a potential future electrical driving.

The resonator consisted of a highly-reflective plane dielectric mirror (R > 99.5% in the range 600-660nm) and a 200-mm radius-of-curvature output coupler (R= 98% @ 600-680 nm). The two mirrors were transparent (T > 90%) at 532 nm. A 17-µm thick film of PMMA ($M_w$ = 9.5×10$^5$ g mol$^{-1}$) doped by 1 wt % of Rhodamine 640 was spin-coated directly onto the plane mirror, allowing 80% of incident radiation at 532 nm to be absorbed in a single pass. The laser cavity was end-pumped (Figure 1) with a pump beam diameter of 140 µm. Two pump sources were used: a "long pulse" Nd:YAG laser (SAGA 230, B.M. Industries Thomson) emitting 7 ns (FWHM) pulses, and a "short-pulse" Nd:YVO$_4$ laser (*PowerChip, Teem Photonics*) with a pulse duration of 0.5 ns. Both sources had 10Hz repetition rate and were frequency-doubled through single-pass in lithium triborate crystal. Laser output energy was monitored with a calibrated photodiode and the emission spectrum detected by a spectrometer (1200 lines/mm grating, 1nm resolution) followed by an Andor technologies DH720 CCD camera.

In the following the thresholds and slope efficiencies are given with respect to the absorbed energy. The laser characteristic curve for long-pulse pumping is shown in Figure 2 (left) for a 4-mm long cavity. A clear lasing threshold is visible at 1.8µJ (11.7mJ/cm$^2$) with maximum output energy of 6 µJ (peak power of 870W). An optical-to-optical efficiency of 43% was measured, corresponding to a power slope efficiency of 52%, or 63% expressed in quantum efficiency. This ranges among the highest reported conversion efficiencies for an organic laser [19] and, to our knowledge, the highest for a thin-film based dye laser . The laser beam was linearly polarized parallel to the pump beam polarization. A typical emission spectrum above threshold is shown in figure 3. It consists of several evenly-spaced peaks, separated by 7.5 nm, which corresponds to the free spectral range of the Fabry-Perot etalon formed by the thick active layer itself. This filtering of the cavity modes by the active layer was verified by spin-coating 10-µm and 5.6-µm-thick films which yielded peaks separated by the expected 13 nm and 22 nm spacings, respectively. Interestingly, tunability could be achieved here over the whole free spectral range of the etalons by taking advantage of the unwanted thickness growth observed at the edges of the mirror, due to the high viscosity of the PMMA solution needed to obtain such thick samples. This corresponded to wavelength tunability over around 20 nm for the 5.6-µm sample. Single-peak (but yet highly multimode) operation was obtained by using a 2.35-µm-thick sample, as shown in the inset of figure 3. However, tunability could not be achieved in this case because of the perfect thickness uniformity of the sample, realized with a lower molar mass PMMA.

Under "short-pulse" pumping (0.5 ns), the laser threshold was lowered to 0.95µJ (6.2 mJ/cm$^2$). The output peak power reached the value of 2 kW for an absorbed pump fluence of 75 mJ/cm$^2$ while in this case maximum output energy was 0.7µJ (fig. 2, right). The conversion and slope efficiencies were 6.3 % and 7 %, respectively. The conversion efficiency is significantly higher than in previously-reported external cavity thin-film dye lasers [17] thanks to the fulfillment of two requirements: the pump intensity is higher than the pump saturation intensity (~MW/cm² for organic dyes) by one or two orders of magnitude for the "long pulse" and "short pulse" pump respectively, which guarantees that a large fraction of the total population can be inverted in a timescale shorter than the radiative lifetime; and the pulse width is long enough to let the laser field build up from spontaneous emission. With the 7-ns pump pulse, steady state is reached within the pulse, as an output energy of 6 µJ emitted from an active area comprising ~$3.10^{12}$ molecules corresponds to an average of 6 photons emitted per molecule and per pulse; with 0.5-ns pumping every molecule emits an average of 0.7 photon/pulse, as oscillation buildup is not over when pumping terminates. A remarkable feature of VECSOLs is that the length of the cavity can be increased up to ~6 cm for "long-pulse" pumping (7ns) and 1 cm for "short-pulse' (0.5 ns) pumping, which enables the insertion of intra-cavity optical components. This is in contrast with previously-reported external-cavity organic semiconductor lasers, where the use of fs pulses for pumping restricted the cavity length to a few microns [17] because of the need for a small oscillation buildup time.

Beam quality was quantified by measuring the $M^2$ factor (figure 4) for both pump sources: the output was diffraction-limited (M²=1). A typical image of the $TEM_{00}$ beam profile is depicted in the inset of figure 4, with two images of other $TEM_{nm}$ Hermite-Gaussian modes that are observable upon fine misalignment of the laser cavity.

Finally, photostability was investigated at fixed absorbed pump fluence above threshold (15.3 mJ/cm$^2$). Under "long-pulse" pumping the output emission decreased to half its initial value after ~ 140000 pulses against ~ 5000 shots with "short-pulse" pumping. This is consistent with a one-order-of-magnitude difference in intracavity peak powers. These values are common for solid-state dye lasers [20] : The plane mirror could be translated without misalignment to maintain stable lasing performance during hours. After the whole sample was degraded, simple rinsing in acetone prior to spin casting yielded a fresh active mirror without loss of performance.

In summary, laser emission from a Vertical External Cavity Surface-emitting Organic Laser (VECSOL) based on PMMA doped with Rhodamine 640 was studied. The efficiency was highly dependent on the pump pulse duration: with a 7-ns pump pulse at 532 nm, output energy (6 µJ), power slope efficiency (52%) and conversion efficiency (43%) were higher than those obtained with 0.5 ns-pulse pumping, for which in contrast higher peak powers (2 kW) were measured. Long-pulse pumping regime allows better photostability and enables lasing with cavity lengths up to 6 cm. The spectrum shows multiple peaks due to the etalon effect of the active layer; tunability could be obtained over up to 20 nm upon playing on thickness variations observed at the edge of the sample. The laser emission was perfectly diffraction-limited ($M^2=1$) for both pump lasers. The VECSOL design is an interesting alternative to traditional low-loss resonators when organic lasers have to meet practical applications requiring a stable, high-quality, and powerful beam.

The authors acknowledge financial support from the ANR (JC/JC call, "BACHELOR" project) and Paris 13 University (BQR credits).

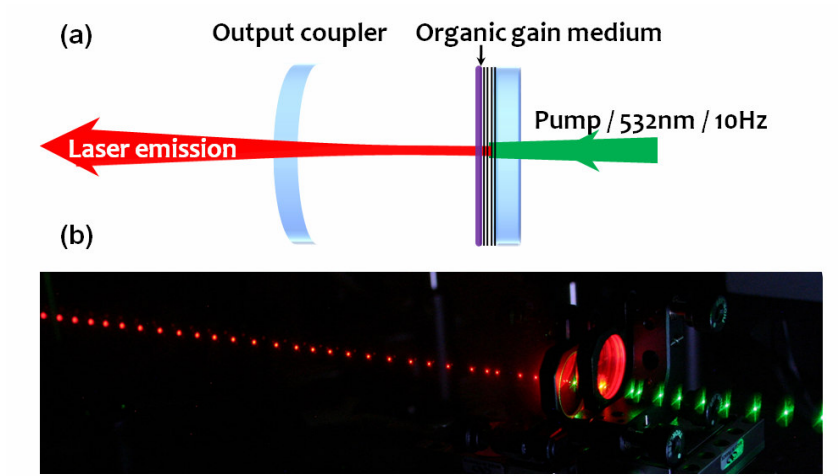

Figure 1 – a) Schematic representation of the plano-concave VECSOL resonator (see text for details). b) Long-exposure time photo of the working VECSOL.

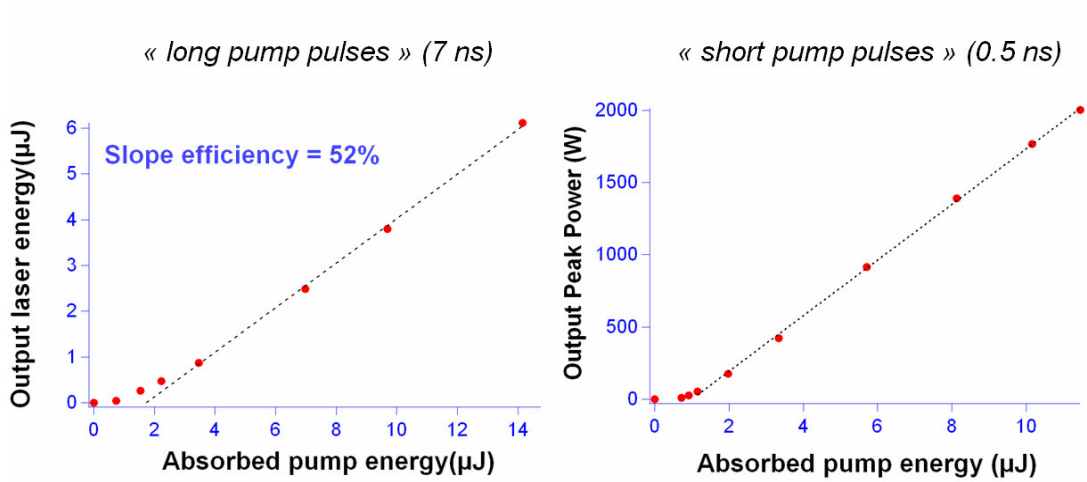

Figure 2 – Output laser energy (left) and output peak power (right) vs. absorbed pump energy under 7 ns and 0.5 ns pumping, respectively. The dashed line is a linear fit to the experimental data.

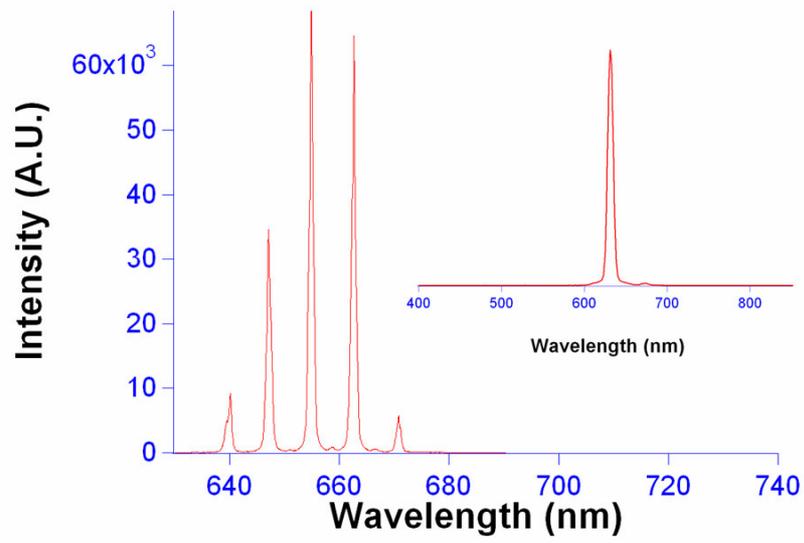

Figure 3– Typical emission spectrum of a 17-µm thick active layer. Inset: single peak laser operation achieved for a 2.35-µm layer.

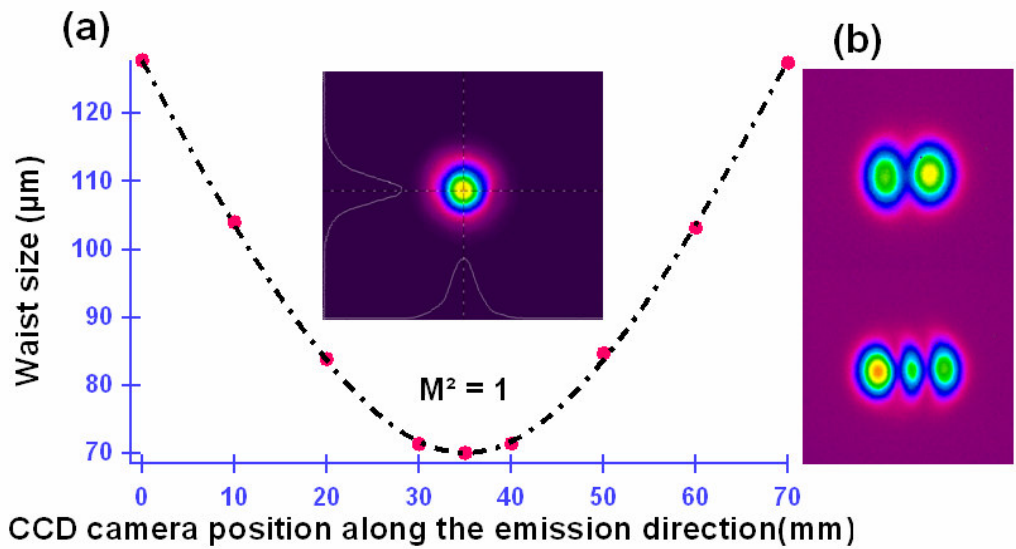

Figure 3 – a) Spatial evolution of the output beam profile along the emission direction. The dotted line is a fit to the experimental data with $M^2=1$. Inset: $TEM_{00}$ laser beam profile. b) $TEM_{10}$ and $TEM_{20}$ images of the VECSOL emission obtained by fine misalignment of the VECSOL cavity.